\documentclass[11pt]{article}
\usepackage[a4paper,margin=2.5cm]{geometry}
\usepackage{graphicx}
\usepackage[colorlinks=true,linkcolor=blue,citecolor=blue,urlcolor=blue]{hyperref}
\usepackage{authblk}
\usepackage{microtype}

\title{\textbf{SpectraFit XPS: a free, browser-based application for X-ray
photoelectron spectroscopy peak fitting and quantification}}
\author[1]{Yongsup Park}
\affil[1]{Department of Physics, Kyung Hee University, Seoul, Republic of Korea\\
\texttt{parky@khu.ac.kr}}
\date{September 2026 \;\textperiodcentered\; describes application version 2.4.6}

\begin{document}
\maketitle

\begin{abstract}
SpectraFit XPS is a free web application for the analysis of X-ray
photoelectron spectroscopy (XPS) data that runs entirely in the browser, with
no installation and no server-side processing: spectra never leave the user's
computer. It combines an interactive Levenberg--Marquardt fitting engine with
true Voigt, Doniach--\v{S}unji\'{c}, and summed Gaussian--Lorentzian line
shapes; Shirley (static and dynamic), linear, and Tougaard (two- and
four-parameter) backgrounds; spin--orbit doublet generation; and
expression-based inter-peak constraints. Quantitative analysis supports ten
relative-sensitivity-factor sets, several kinetic-energy correction schemes
including TPP-2M inelastic mean free paths, angular correction, and atomic and
weight percentages, with the underlying database and algorithms adapted from
the open-source KherveFitting project. The application reads and writes
ISO~14976 (VAMAS) files---including multi-region and depth-profile data and
load-time transmission-function correction---and imports common vendor and
spreadsheet formats. A publication plotter exports journal-ready SVG/PNG
figures. The quantification pipeline reproduces KherveFitting reference values
exactly in a 30-case regression suite, part of an automated test set of 100
tests. The application, an illustrated bilingual user guide, and a synthetic
demonstration dataset are freely available at
\url{https://spectrafit-xps.web.app/}.
\end{abstract}

\section{Introduction}
X-ray photoelectron spectroscopy is a standard tool for surface chemical
analysis, and curve fitting of core-level spectra remains the step where most
interpretive decisions are made: choice of background, line shape, number of
components, constraints between them, and sensitivity factors for
quantification~\cite{Briggs2003,Fairley2021}. The software landscape for this
step is dominated by commercial desktop packages and instrument-vendor tools;
capable open-source alternatives exist---notably
KherveFitting~\cite{Kherve}, a Python desktop application---but desktop
software of any kind carries an installation barrier that is surprisingly
consequential in practice: students on managed laboratory computers, quick
looks at data on a different machine, teaching settings, and collaborators who
only occasionally touch XPS data.

\begin{figure}[tb]
\centering
\includegraphics[width=\linewidth]{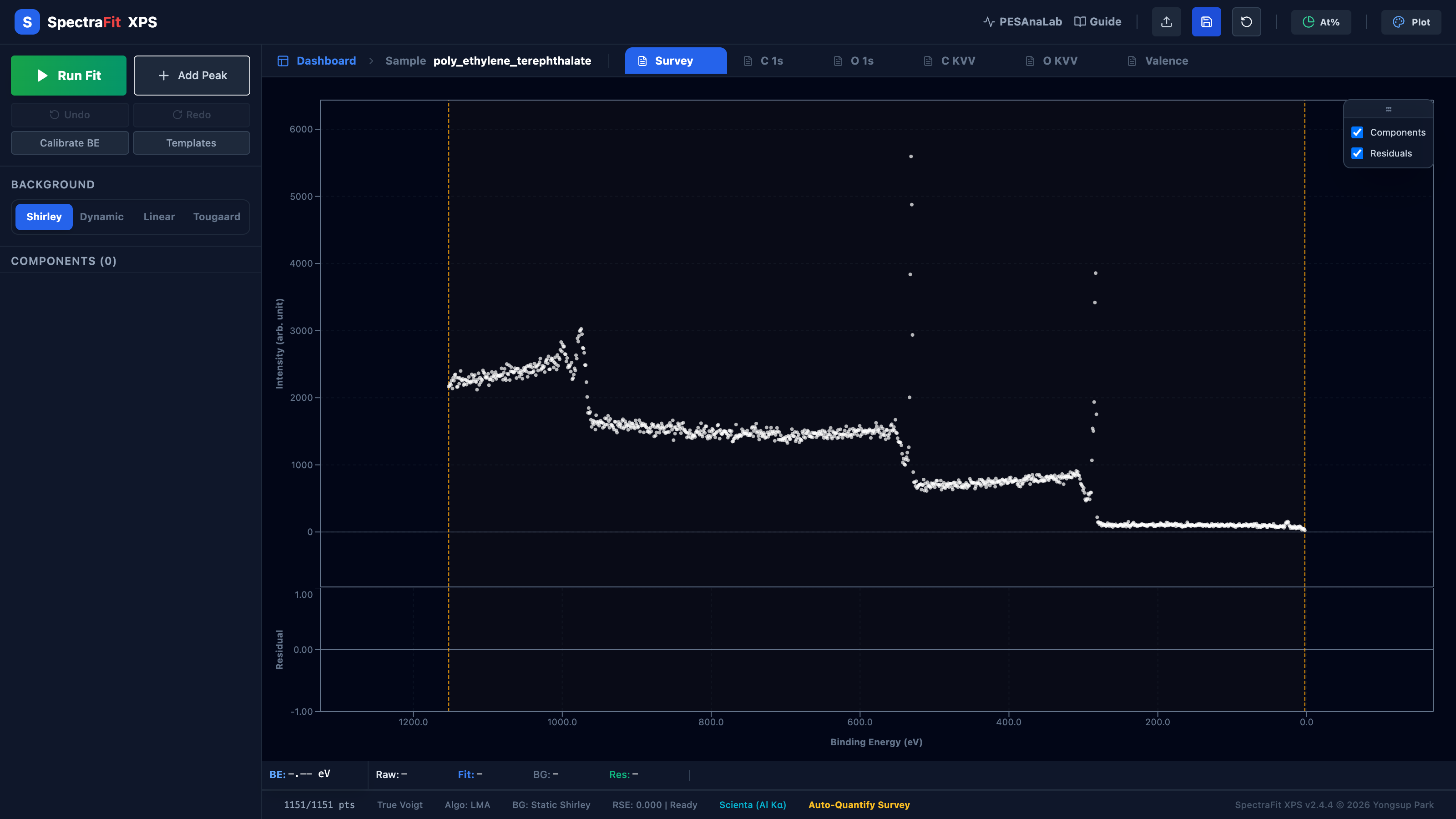}
\caption{The SpectraFit XPS workspace with a poly(ethylene terephthalate)
survey spectrum loaded from a VAMAS file~\cite{BeamsonBriggs}. Left: fitting
controls (background selector and component list). Center: spectrum with
region-of-interest handles and, below, the residual panel. The footer reports
point counts, the active line-shape and background models, the residual
standard error, and the instrument auto-detected from the file header.}
\label{fig:workspace}
\end{figure}

SpectraFit XPS approaches this gap from the browser. It is a single-page web
application that performs every computation client-side, so that (i) nothing
needs to be installed or licensed, (ii) the tool runs identically on any
operating system with a modern browser, and (iii) measured data are never
transmitted anywhere---a property that matters for unpublished results and
industrial samples. The application is aimed at the everyday fitting workflow:
load a file, bracket a region, choose a background, place peaks, fit, quantify,
and export a publication-quality figure (Fig.~\ref{fig:workspace}).

This note describes the physics content and the design of the application at
version 2.4.6, its validation against established references, and its
limitations.

\section{Design overview}
The application is written in TypeScript with React and is served as static
files from Firebase Hosting; there is no application server. All numerics run
in the browser's JavaScript engine. Interactive responsiveness is treated as a
design constraint: backgrounds and residuals update in real time while peaks
are dragged, and the Levenberg--Marquardt loop yields intermediate states so
that fits animate rather than block the interface.

Sessions are organized as \emph{samples} containing \emph{regions} (e.g.\ a
survey plus core-level windows), each with its own background model, peaks, and
fit range; depth-profile files open as slice series navigable with a slider. A
complete session can be saved to a single JSON project file and restored later.
An illustrated user guide (English and Korean) is available from the
application header and at \url{https://spectrafit-xps.web.app/guide}, and a
bundled, fully synthetic PET-like example dataset lets a first-time visitor
try region fitting and quantification without their own data.

\section{Fitting engine}
\subsection{Line shapes}
Three line-shape families are provided per component:
\begin{itemize}
\item \textbf{True Voigt} --- a numerical convolution of Gaussian and
Lorentzian profiles with independently adjustable Gaussian and Lorentzian
widths, rather than the pseudo-Voigt linear combination often used for speed.
\item \textbf{Doniach--\v{S}unji\'{c}} --- the asymmetric line shape of
metallic core levels~\cite{DoniachSunjic1970}, with the asymmetry parameter
fittable or lockable per component.
\item \textbf{Summed Gaussian--Lorentzian (SGL)} --- the
Gaussian--Lorentzian sum parameterization familiar from CasaXPS-style
workflows~\cite{Fairley2021}; the mixing ratio is held fixed during
optimization while position, height, and width are refined, matching common
practice.
\end{itemize}
Spin--orbit doublets are generated from a built-in table of orbital rules
(area ratios 2:1, 3:2, 4:3 for $p$, $d$, $f$ levels and characteristic
splittings); the splitting and intensity ratio can each be locked or released
as fit parameters.

\subsection{Backgrounds}
Four background models are available per region: static
Shirley~\cite{Shirley1972}; a dynamic Shirley variant whose parameters are
optimized together with the peaks; linear; and Tougaard backgrounds using the
two-parameter universal cross-section as well as the four-parameter (U4)
form~\cite{Tougaard1988,Tougaard1997}, with a one-click universal preset
($B=2866$~eV$^2$, $C=1643$~eV$^2$).

\subsection{Optimization and constraints}
Model parameters are refined by a Levenberg--Marquardt
loop~\cite{Levenberg1944,Marquardt1963} with early termination when an
iteration improves the residual sum of squares by less than $10^{-6}$. The
residual panel plots measured-minus-model at the measured points only, and the
footer reports the residual standard error. Components can be linked to the
first (``master'') peak by width, relative position, or height ratio, and
version 2.1 added expression-based constraints (e.g.\
\texttt{P1.pos + 1.18} or \texttt{P1.fwhm\_g * 1.2}) that are re-evaluated at
every iteration. One-click binding-energy calibration against C~1s, Au
4f$_{7/2}$, Ag 3d$_{5/2}$, or Cu 2p$_{3/2}$ references can be applied to a
single region or to all regions of a sample, and multi-peak models can be
stored as reusable templates.

\begin{figure}[tb]
\centering
\includegraphics[width=\linewidth]{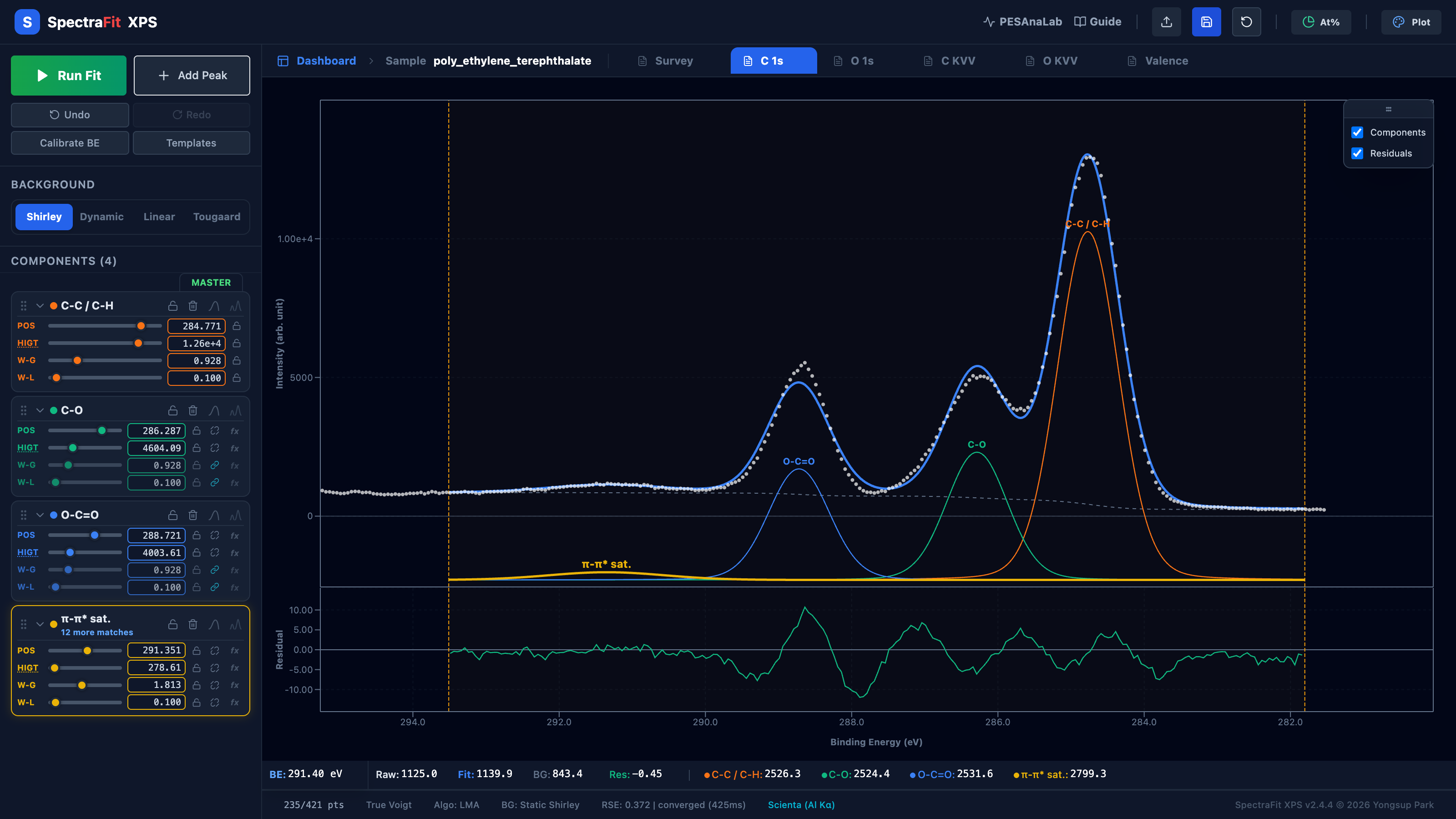}
\caption{The region-fitting view during a four-component fit of the PET
C~1s region (data from the XPS of Polymers database~\cite{BeamsonBriggs}):
C--C/C--H, C--O, and O--C=O components with a $\pi\to\pi^{*}$ shake-up
satellite on a Shirley background. Each component card exposes
per-parameter lock, master-link, and expression controls; dashed handles
delimit the fitted range; the lower panel shows the live residual, and the
footer reports the line shape, optimizer, background, residual standard
error, and convergence time.}
\label{fig:fitwin}
\end{figure}

\begin{figure}[tb]
\centering
\includegraphics[width=0.8\linewidth]{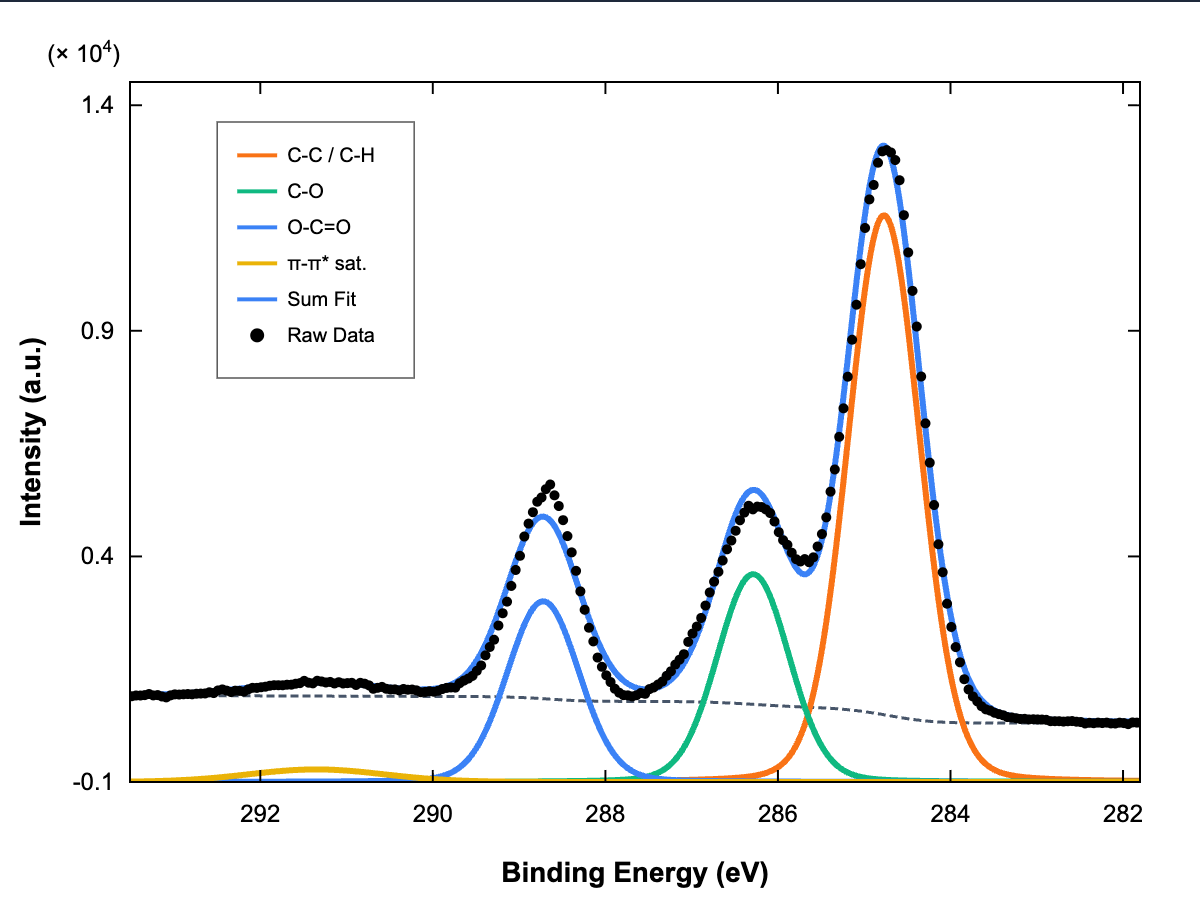}
\caption{Publication-plotter export of the fit shown in
Fig.~\ref{fig:fitwin}, rendered as a white-background, print-ready figure
with configurable axes, fonts, and legend.}
\label{fig:petfit}
\end{figure}

Figure~\ref{fig:fitwin} shows this machinery in the view where most analysis
time is spent: per-component parameter cards with lock, link, and expression
controls, draggable range handles, a live residual panel, and a footer
reporting the optimizer state. The publication plotter renders the same fit
as an export-ready figure (Fig.~\ref{fig:petfit}).

\section{Quantification}
\begin{figure}[tb]
\centering
\includegraphics[width=\linewidth]{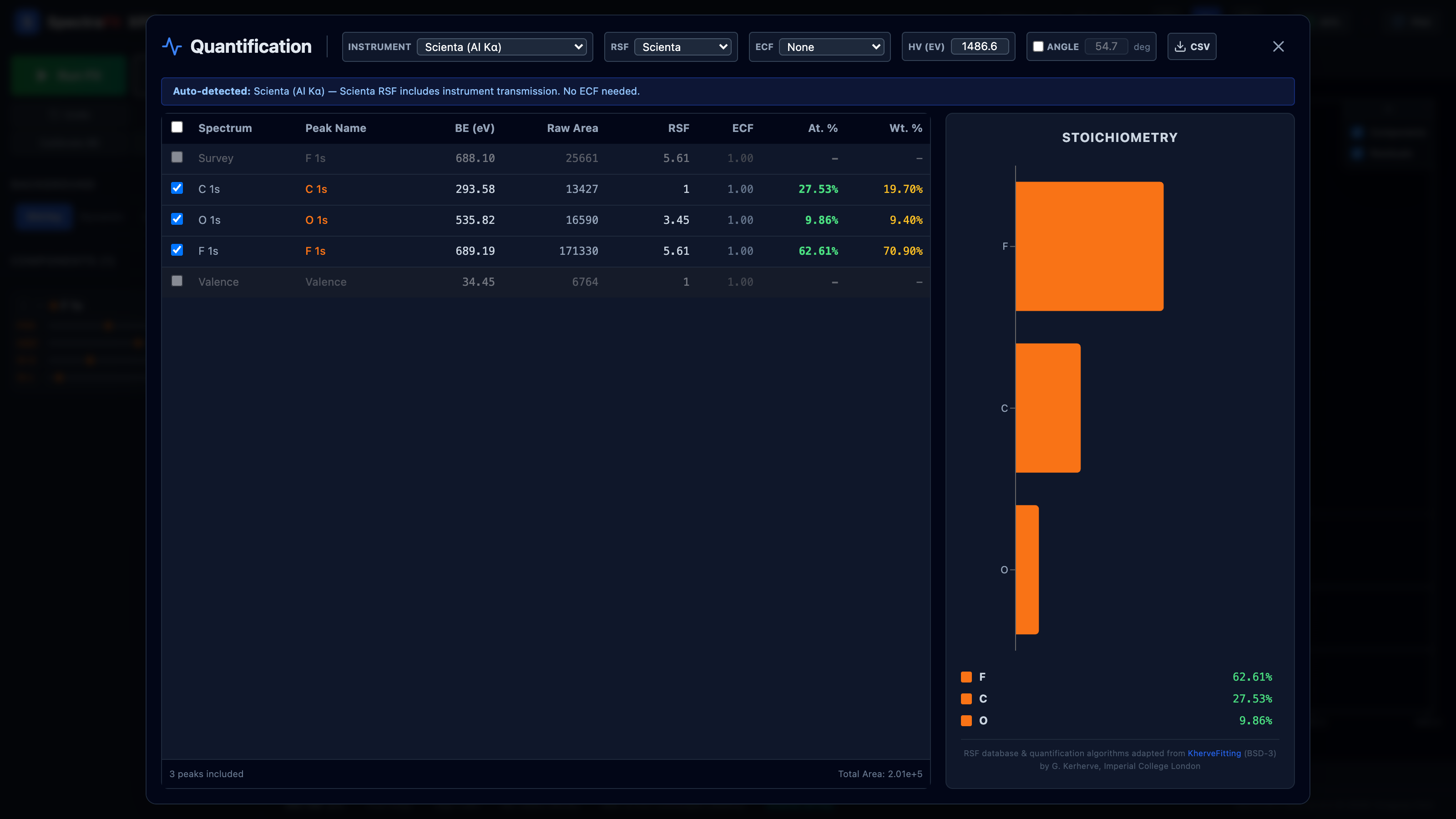}
\caption{Quantification of Fomblin~Y (data from~\cite{BeamsonBriggs}). The
spectrometer was auto-detected from the VAMAS header (banner), configuring the
RSF set and corrections; non-core-level rows (survey, valence) are excluded by
checkbox. The result, C/O/F = 27.5/9.9/62.6~at.\%, reflects the expected
fluorine-dominated composition.}
\label{fig:quant}
\end{figure}

Atomic percentages are computed from fitted (or automatically detected) peak
areas $A_i$ as $x_i = (A_i/S_i F_i L_i)/\sum_j (A_j/S_j F_j L_j)$, where $S_i$
are relative sensitivity factors, $F_i$ an optional kinetic-energy correction,
and $L_i$ an optional angular correction; weight percentages follow from
standard atomic masses. The implementation adapts the database and pipeline of
the open-source KherveFitting project~\cite{Kherve} (BSD-3 license):
\begin{itemize}
\item \textbf{RSF sets} --- Wagner~\cite{Wagner1981} and
Scofield~\cite{Scofield1976} (Al and Mg anodes) plus instrument-specific sets
(Thermo, Kratos, Scienta, PHI, Shimadzu).
\item \textbf{Kinetic-energy corrections} --- none, $KE^{0.6}$,
$KE^{1.0}$, effective attenuation lengths after Seah~\cite{Seah2001}, or
TPP-2M inelastic mean free paths~\cite{TPP2M} with average matrix parameters.
\item \textbf{Angular correction} --- an $L(\beta,\theta)$ factor for
non-magic-angle geometries (default $54.7^{\circ}$).
\end{itemize}
When a VAMAS file identifies the spectrometer, the application selects a
matching preset (RSF set, correction scheme, photon energy) automatically and
warns about double-correcting combinations, e.g.\ a manufacturer RSF set that
already contains the transmission function combined with a further
transmission-like correction. Figure~\ref{fig:quant} shows the quantification
panel for the perfluoropolyether Fomblin~Y: with the auto-selected Scienta
preset, the fitted C~1s, O~1s, and F~1s regions give C/O/F =
27.5/9.9/62.6~at.\%, consistent with the fluorine-dominated stoichiometry of
the material; the fluorine-shifted C~1s at $\approx 293.6$~eV is correctly
labeled from its region metadata rather than mis-assigned by binding energy.

A one-click survey quantification (peak detection, element identification, and
approximate atomic percentages without fitting) is provided for wide scans and
is deliberately conservative---identification is restricted to commonly
encountered elements via their core photoelectron lines, and explicit region
labels are trusted. For a PET survey it yields C:O $\approx 70{:}30$, matching
the C$_{10}$H$_8$O$_4$ repeat unit. Its limitations are documented (see
Sec.~\ref{sec:limits}).

\section{Data formats and interoperability}
The application reads ISO~14976 VAMAS files~\cite{Dench1988} with automatic
region splitting, kinetic-to-binding-energy conversion, depth-profile slice
grouping via experiment variables, and load-time transmission-function
correction when the file carries one; it also \emph{writes} VAMAS, so
processed spectra can move to CasaXPS, KherveFitting, MultiPak, or any other
ISO-14976 reader. Additional import formats are Excel/CSV/tab-separated text
(including multi-column vendor exports such as Thermo Avantage), XPS ASCII
(\texttt{.asc}), Physical Electronics MRS, and VG-Microtech fixed-format
files. Fits export as spreadsheet or column text including every component
curve, and the publication plotter produces white-background SVG or PNG
figures with adjustable axes, fonts, legend, and annotations
(Fig.~\ref{fig:petfit}).

\section{Validation}
The numerical core is covered by an automated test suite (100 tests at version
2.4.6) spanning line-shape evaluation, background construction, parsers for
each import format, a VAMAS writer round trip, element identification, and
quantification. The quantification pipeline is verified against KherveFitting
reference calculations in a 30-case regression suite with zero deviation, and
the project's quality gate requires agreement within $\pm 2\%$ for any change
touching the pipeline. Example analyses on database polymer
spectra~\cite{BeamsonBriggs} reproduce textbook results: the PET C~1s region
resolves into the expected three chemical states plus shake-up satellite
(Fig.~\ref{fig:petfit}), and Fomblin~Y quantification returns the expected
composition (Fig.~\ref{fig:quant}).

\section{Privacy and availability}
All parsing, fitting, and quantification run locally; spectra are never
uploaded. The application collects only anonymous feature-usage statistics
(e.g.\ that a file of a given format was imported) with IP anonymization, and
an optional Google Drive integration exists solely for users who wish to open
or save their own files from cloud storage on devices without local storage.
SpectraFit XPS is free to use at \url{https://spectrafit-xps.web.app/}; the
illustrated user guide is served at \texttt{/guide} (English) and
\texttt{/guide-ko} (Korean). Algorithms and data adapted from
KherveFitting~\cite{Kherve} are used under the BSD-3-Clause license with
attribution in the application and documentation.

\section{Limitations and outlook}\label{sec:limits}
The automatic survey quantification is approximate and currently unreliable
for complex multi-element samples: weak transition-metal peaks beneath strong
C/O contamination can be missed, and spurious elements can be reported. This
is stated in the release notes, and region-by-region fitting---where element
labels are reliable---is recommended for quantitative work; a guided
element-selection mode is planned. Uncertainty estimates for fitted parameters
are not yet reported. Planned directions include valence-band and
UPS/inverse-photoemission analysis and broader native instrument-format
support.

\section*{Acknowledgments}
The RSF database, Tougaard background, TPP-2M implementation, and
quantification pipeline are adapted from KherveFitting by Gwilherm Kerherve
(Imperial College London), whose open-source release made this work
possible~\cite{Kherve}. Polymer reference spectra used in the illustrated
examples are from the XPS of Polymers database~\cite{BeamsonBriggs}.

\end{document}